# DEEP LEARNING-BASED CLASSIFICATION OF BREAST CANCER MOLECULAR SUBTYPES FROM H&E WHOLE-SLIDE IMAGES

Masoud Tafavvoghi[a,*], Anders Sildnes[b], Mehrdad Rakaee[c,e,f], Nikita Shvetsov[b], Lars Ailo Bongo[b], Lill-Tove Rasmussen Busund[c,d] and Kajsa Møllersen[a]

[a] Department of Community Medicine, Uit The Arctic University of Norway, Tromsø, Norway
[b] Department of Computer Science, Uit The Arctic University of Norway, Tromsø, Norway
[c] Department of Medical Biology, Uit The Arctic University of Norway, Tromsø, Norway
[d] Department of Clinical Pathology, University Hospital of North Norway, Tromsø, Norway
[e] Department of Medicine, Brigham and Women's Hospital, Harvard Medical School, Boston, USA
[f] Department of Cancer Genetics, Oslo University Hospital, Oslo, Norway

**Abstract**
Classifying breast cancer molecular subtypes is crucial for tailoring treatment strategies. While immunohistochemistry (IHC) and gene expression profiling are standard methods for molecular subtyping, IHC can be subjective, and gene profiling is costly and not widely accessible in many regions. Previous approaches have highlighted the potential application of deep learning models on H&E-stained whole slide images (WSI) for molecular subtyping, but these efforts vary in their methods, datasets, and reported performance. In this work, we investigated whether H&E-stained WSIs could be solely leveraged to predict breast cancer molecular subtypes (luminal A, B, HER2-enriched, and Basal). We used 1,433 WSIs of breast cancer in a two-step pipeline: first, classifying tumor and non-tumor tiles to use only the tumor regions for molecular subtyping; and second, employing a One-vs-Rest (OvR) strategy to train four binary OvR classifiers and aggregating their results using an eXtreme Gradient Boosting (XGBoost) model. The pipeline was tested on 221 hold-out WSIs, achieving an overall macro F1 score of 0.95 for tumor detection and 0.73 for molecular subtyping. Our findings suggest that, with further validation, supervised deep learning models could serve as supportive tools for molecular subtyping in breast cancer. Our codes are made available to facilitate ongoing research and development.

**Introduction**
Breast cancer accounts for 12.5% of all diagnosed cancer types globally, with around 2.3 million new cases and 685,000 fatalities annually, and is expected to grow to 3 million newly diagnosed cases and 1 million deaths by 2040 [1]. Breast cancer is a heterogeneous disease, and its outcome depends on patients' demographic factors and tumor characteristics, including the crucial distinction among molecular subtypes, which play a significant role in determining treatment strategies. Broadly, breast cancer has four molecular subtypes: luminal A (LumA), luminal B (LumB), HER2-enriched (HER2), and basal-like (BL). Normally, BL tumors exhibit higher rates of recurrence during the initial five years following detection and treatment, but they show higher response to chemotherapy. On the other hand, luminal cancers, accounting for 60-70% of all breast cancers [2], respond poorly to chemotherapy, and LumA tumors have lower early recurrence compared to other breast cancer molecular subtypes [3, 4]. Therefore, identifying the molecular subtypes of breast cancer is crucial for treatment decisions.

Currently, gene expression profiling serves as a new technology for breast cancer molecular subtyping, which is substantially more expensive and not available in all healthcare systems [5]. As a result, immunohistochemistry (IHC) staining is still widely used to classify the subtypes in clinical practice. IHC staining involves using specific antibodies to detect and visualize specific proteins' presence, localization, and abundance within breast cancer tissue samples. The IHC staining is typically performed for four key biomarkers: Estrogen Receptor (ER), Progesterone Receptor (PR), Human Epidermal Growth Factor Receptor 2 (HER2), and antigen Ki-67. Based on the results of these four stainings, breast cancer can be classified into four main molecular subtypes: LumA, LumB, HER2, and BL [6].

In recent years, deep learning has emerged as a transformative technology in many fields, notably in medical image analysis [7, 8]. As a branch of machine learning, deep learning uses neural networks with multiple layers to identify complex patterns in raw data without manual feature extraction, a major improvement over traditional methods. This capability is especially valuable in medical imaging, where it processes large volumes of data to deliver precise and automated analyses. Deep learning has demonstrated exceptional performance in histopathology analysis, a critical aspect of cancer diagnosis and research. It can accurately identify subtle features in histopathological slides, such as cell morphology, tissue structures, and biomarker expressions, with high precision [9–11]. Such breakthroughs not only streamline the diagnostic process but also hold the potential to improve the accuracy and reproducibility of results, ultimately benefiting patient care and advancing our understanding of complex diseases like breast cancer.

Traditional methods for classifying breast cancer molecular subtypes rely heavily on histopathological examination, which is often time-consuming, subjective, and sometimes inconsistent in interpretation [12]. Additionally, new technologies, while promising, are indeed expensive and may not be readily available in many countries and healthcare systems. In contrast, Hematoxylin and Eosin (H&E) staining, as the gold standard in histopathology [13], offers a more accessible and cost-effective approach. This technique is widely recognized for its reliability in tissue characterization and pathology diagnosis.

In the field of digital pathology, several studies have utilized H&E-stained histopathological images for classifying breast cancer molecular subtypes, each facing certain limitations. Couture et al. [14] and Jaber et al. [15] demonstrated the potential of deep learning for predicting BL and non-BL subtypes, highlighting its significant clinical implications, especially in resource-limited settings. However, their models faced challenges such as misclassification risks due to subtype heterogeneity and the presence of non-cancerous tiles (also known as patches) in cancer-rich clusters. In a similar vein, Abbasi et al. [16] noted the variability in model performance across different scanners, underscoring the necessity for equipment-specific model tuning. Liu et al. [17] explored the use of weakly supervised learning models on a private image dataset to classify breast cancer subtypes, concluding that while AI can aid preliminary screening, it cannot yet fully replace traditional human analysis.

Expanding on these aforementioned studies, we hypothesize that H&E-stained histopathology images contain sufficient information to classify breast cancer molecular subtypes by exhibiting different morphological patterns in the breast tissue. To evaluate this hypothesis, we have developed a multi-stage model that labels H&E WSIs after performing non-relevant tile exclusion, color normalization, and tile classification. Additionally, we enriched our analysis with an expanded dataset combining slides from publicly available datasets, enhancing the reproducibility of our results.

Building on this hypothesis, this study aims to investigate whether H&E-stained histopathology images contain sufficient information to classify molecular subtypes without carrying out additional analyses like IHC staining or gene expression profiling. This approach can offer cost and time efficiency, simpler diagnostics, wider accessibility, reduced high-tech reliance, better resource use, and suitability for large-scale research.

**Methods**
The procedure of classification of breast cancer molecular subtypes in this paper consists of two main parts. In the first part, we trained a deep learning model for classifying tumor and non-tumor tiles in a WSI, intending to utilize only tumor regions for the classification of molecular subtypes. In the second part, we trained separate classifiers for breast cancer molecular subtyping. In the following subsections, we delve into descriptions of utilized datasets, preprocessing steps, and training of our models. The CNN classifier codes used in this study were adapted from the repository by Foersch et al. [18] and modified as needed. The codes are available at https://github.com/uit-hdl/BC_MolSubtyping.

**Datasets**

**Classification of tumor and non-tumor regions**

In this study, we used two publicly available sets of data, one for classifying tumor and non-tumor tiles in H&E-stained WSIs and the other set for the classification of breast cancer molecular subtypes (Table 1). To develop and test a binary deep learning model for tumor and non-tumor classification of image tiles, we used 195 WSIs of TCGA-BRCA [19] and 129 WSIs of BRACS [20] datasets. The BRACS dataset has annotated regions of interest that can be used to extract tumor tiles from WSIs. The TCGA-BRCA dataset has no officially published annotations of regions in WSIs. However, there are 195 WSIs from TCGA-BRCA in the DRYAD [21] dataset that have annotated tumor regions. Since both images and annotations in the DRYAD dataset are downsized to a 1/10 scale, we replicated the annotated regions for the full-size WSIs and then extracted image tiles from those areas.

**Table 1:** Characteristics of breast H&E WSIs used in this study

| Classification of breast tumor and non-tumor image tiles | | | | | |
|---|---|---|---|---|---|
| Dataset | #WSIs | Source | Pixel size [$\mu$m/pixel] | Scanner | Image format |
| TCGA-BRCA | 195 | USA | 0.25, 0.50 | Variant | .svs |
| BRACS | 129 | Italy | 0.25, 0.50 | Aperio AT2 | .svs |
| Classification of breast cancer molecular subtypes | | | | | |
| Dataset | #WSIs | Source | Pixel size [$\mu$m/pixel] | Scanner | Image format |
| TCGA-BRCA | 980 | USA | 0.25, 0.50 | Variant | .svs |
| CPTAC-BRCA | 382 | USA | 0.25, 0.50 | Variant | .svs |
| HER2-Warwick | 71 | UK | 0.23 | NanoZoomer C9600 | .ndpi |

**Classification of breast cancer molecular subtypes**

The selection of datasets for classifying breast cancer molecular subtypes was based on the availability of pertinent labels for WSIs. According to a review paper we published earlier on publicly available datasets of breast cancer histopathology images [22], there were only two datasets featuring these specific labels: TCGA-BRCA and CPTAC-BRCA [23]. From 1,134 available WSIs in the TCGA-BRCA dataset, we acquired 980 WSIs labeled with molecular subtypes and excluded 154 WSIs that either lacked labels or were categorized as "Normal-like" tumors. In the CPTAC-BRCA dataset, there are 640 WSIs of breast tissue. Of these, 382 WSIs are labeled with molecular subtypes, and the remaining 258 -either unlabeled or labeled as "Normal-like"- were excluded from this study. In both TCGA-BRCA and CPTAC-BRCA datasets, the major and minor classes were LumA and HER2, with an overall share of 50.1% and 8.6%, respectively, which can cause significant dataset imbalance. To mitigate this imbalanced distribution of classes, we added the HER2-Warwick dataset [24], which has 86 H&E-stained WSIs of invasive breast carcinomas from 86 patients, 71 of those with positive HER2 expression scores that were used in our study (Table 1). Detailed distribution of breast molecular subtypes in each dataset is presented in Table 2.

**Table 2:** Distribution of WSIs in the four classes used for the classification of breast molecular subtypes

| Dataset | LumA | LumB | HER2 | Basal | Excluded |
|---|---|---|---|---|---|
| TCGA-BRCA | 507 | 219 | 78 | 176 | 122 unlabeled and 37 normal-like WSIs |
| CPTAC-BRCA | 176 | 44 | 39 | 123 | 258 unlabeled and 13 normal-like WSIs |
| HER2-Warwick | 0 | 0 | 71 | 0 | 15 WSIs without positive HER2 expression scores |
| Total | 683 | 263 | 188 | 299 | 445 |

**Preprocessing**

**Classification of tumor and non-tumor regions**

Typically, WSIs are extremely large, containing billions of pixels, and cannot be directly fed into any deep learning model. Therefore, regions of interest within the WSIs are divided into tiles to be compatible with these algorithms. For the classification of tumor and non-tumor tiles, we used QuPath [25] software to make and extract non-overlapping tiles with size 512×512 at 0.5 μm/pixel magnification from the annotated tumor regions in TCGA-BRCA and BRACS WSIs. This yielded 38,392 tumor tiles from the TCGA-BRCA and BRACS datasets, sufficient to serve as the tumor class for training a binary classifier. Additionally, we extracted 37,407 tiles from the non-tumor areas in the same WSIs to have roughly balanced tumor and non-tumor classes. The non-tumor class comprised normal tissues, folded tissue areas, marker signs, and white areas on the slides to prevent extra steps to remove low-quality and white tiles for the classification of breast molecular subtypes in the subsequent steps (Fig. S1). All 75,799 tiles were then split into 70%, 15%, and 15% for training, validation, and testing, respectively, ensuring that tiles from each WSI were assigned exclusively to one of these sets. This split also maintained the balance of classes in each set. Fig. 1 illustrates the workflow of preprocessing steps for classifying breast H&E images into tumor and non-tumor classes.

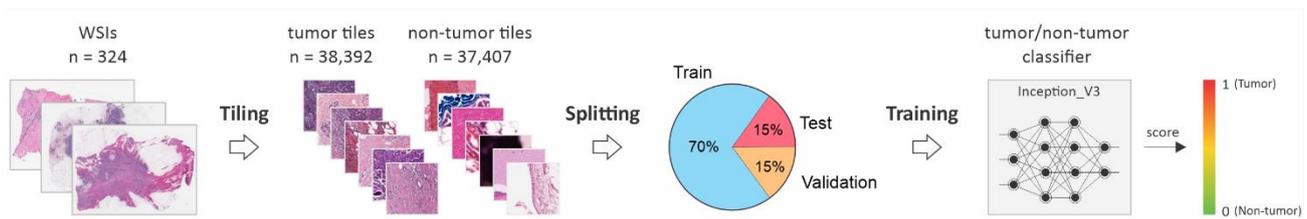

**Figure 1**: The workflow of training a deep binary classifier for tumor and non-tumor breast tiles.

**Classification of breast cancer molecular subtypes**

Data preprocessing for the classification of breast cancer molecular subtypes was more comprehensive. In the first step, we used QuPath to detect the tissue areas in 1,433 WSIs. Detected areas were then divided into tiles with size 512×512 at 0.5 μm/pixel magnification in TIFF format without compressing the image data. For LumA, LumB, and BL classes, we extracted the tiles without overlapping. However, to increase the number of extracted tiles (instances) for the minority class, we set an overlap of 64 pixels for the HER2 WSIs. This is due to the fact that WSIs in the HER2-Warwick dataset originate from biopsies rather than tissue resections, resulting in fewer image tiles compared to surgical resections (Fig. S2).

Since normal areas and artifacts in the image do not contribute to our classification task, we chose to use only tumor tiles. However, most of the WSIs in the TCGA-BRCA dataset and all WSIs of the CPTAC-BRCA and HER2-Warwick datasets lacked annotations of tumor areas. Therefore, to take only tumor tiles, we fed all 3,571,651 extracted tiles to the earlier trained binary tumor/non-tumor classifier to determine the likelihood of each tile belonging to the tumor class. Following that, to create a balanced dataset with a nearly equal number of tiles in each of the four classes of breast cancer molecular subtypes, namely LumA, LumB, HER2, and BL, we used the minor class (HER2) as the reference class with 278,675 tumor tiles and balanced the four breast cancer classes based on that. Since the number of WSIs in each class was different, we took 441, 1180, and 1410 random tumor tiles per WSI from LumA, LumB, and BL classes, respectively, to have a roughly equal distribution of tiles among each class (Fig. 2). For a detailed illustration of data partitioning, see Fig. S3. It is important to note that the actual counts of selected tumor tiles for model development differ from the expected values. This discrepancy arose because many WSIs contained small tumor regions, resulting in fewer tumor tiles than specified for each class.

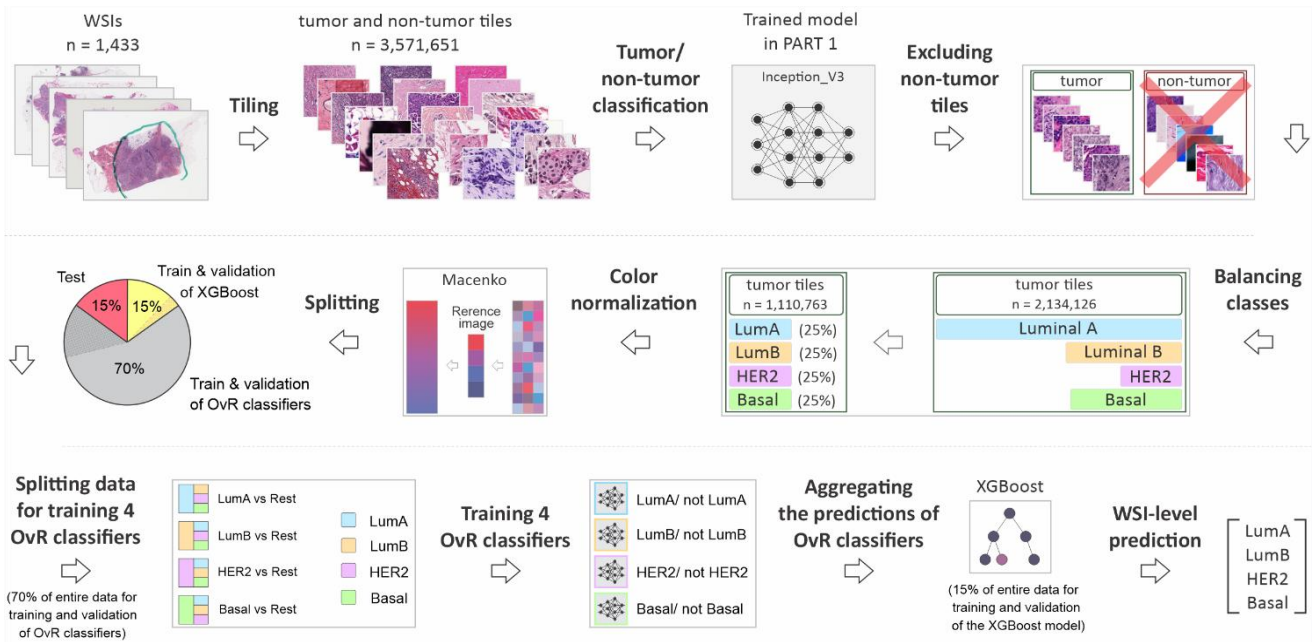

**Figure 2**: The workflow of classifying breast H&E WSIs into four molecular subtypes, where the results of four binary OvR classifiers were aggregated by an XGBoost model to predict the subtype in WSIs.

The selected tumor tiles were then color normalized using the Macenko [26] method to ensure consistency in color representation across all images and minimize potential variations caused by differences in image acquisition conditions. To normalize the color of tumor tiles, we generated a reference image using Python by picking one random tile out of the tumor tiles per 256 randomly selected WSIs (Fig. S4). Following that, to mitigate the risk of overfitting, the color-normalized tumor tiles were divided into three distinct sets: 70% for training and validation of convolutional neural networks (CNNs), 15% for training and validation of the eXtreme Gradient Boosting (XGBoost) [27] model, and the remaining 15% for testing the entire pipeline on the classification of BC molecular subtypes. Each set maintained a stratified split to ensure a proportional representation of images per class, and WSIs from each patient were restricted to a single set, ensuring that different images from the same patient were not included in multiple sets.

**Model training**
**Deep convolutional neural networks**
**Classification of tumor and non-tumor regions**
To train the tumor/non-tumor classifier, we used the Inception_V3 [28] architecture with pre-trained weights, implemented in PyTorch (version 1.7.1 + cu110). Inception_V3 is known for its efficiency in capturing complex hierarchical features through the use of Inception modules, which perform convolutions of various sizes (1x1, 3x3, 5x5) within the same layer, allowing the network to capture multi-scale features effectively. The architecture also incorporates auxiliary classifiers at intermediate layers to help propagate gradients and improve convergence during training. Inception_V3's modular approach allows it to capture intricate patterns and features within histopathology images [29, 30], making it a suitable choice for this classification task. The auxiliary classifiers embedded within the network also aid in preventing gradient vanishing issues, which can be prevalent in deep networks.

The training was performed using an RTX-3090 GPU with 24 GB of VRAM. The hyperparameters were set as follows: a batch size of 64, a learning rate of 1e-5, and a dropout rate of 0.33 to prevent overfitting. We used the ADAM optimizer, known for its robustness to sparse gradients, and the cross-entropy loss function, which is suitable for binary classification.

**Classification of breast cancer molecular subtypes**

For classifying breast cancer molecular subtypes from H&E-stained images, we trained four separate binary classifiers using the One-vs-Rest (OvR) strategy to simplify a complex multi-class classification into binary tasks. For training and validation, 70% of the selected tumor tiles were allocated (the gray part of the pie chart in Fig. 2), with 80% (56% of the total data in light gray) for training and 20% (14% of the total data in dark gray) for validation (model fine-tuning). Each classifier was trained using the ResNet-18 architecture [31], with pre-trained weights, incorporating all tiles from the target subtype and one-third from each of the other three subtypes as the rest class, ensuring balanced binary classes (Fig. 2 and Fig. S3) for training the models.

This architecture was chosen due to its proven effectiveness in various image classification tasks, offering a good balance between depth and computational efficiency [32, 33]. Moreover, ResNet-18's relatively shallow depth compared to deeper variants ensures faster training and inference times without significant loss in accuracy, which is crucial when training four deep classifiers with large-scale datasets. The training was performed using the same hardware with the following hyperparameters: a batch size of 128, a learning rate of 5e-6, a dropout rate of 0.33, the ADAM optimizer, and the cross-entropy loss function.

**Thresholding**

The output of a binary classifier for the target class is a single score between 0 and 1 for each image tile. To aggregate these scores into a definitive classification for each WSI, we set a threshold for each classifier, which determines the predicted class of each tile. Using a fixed threshold of 0.5 for assigning classes to images can be sub-optimal [34, 35]. Therefore, we employed Precision-Recall (PR) analysis to establish the optimum decision thresholds for each classifier. PR analysis aids in identifying a threshold that balances precision (positive predictive value) and recall (sensitivity) effectively. The PR curve plots precision against recall, enabling the selection of a threshold that maximizes the classifier's performance. Although our classes are balanced, this approach is particularly relevant as we train four separate binary OvR classifiers, where the target class in each classifier is treated as the positive class. This allows us to fine-tune each OvR classifier for optimal performance in distinguishing the target class from all others. We used only the validation set, which constitutes 20% of the classification data (equivalent to 14% of the total tumor tiles), to adjust the threshold of OvR classifiers. The optimum thresholds for LumA, LumB, HER2, and BL classifiers were determined to be 0.434, 0.415, 0.481, and 0.424, respectively. These thresholds were applied in subsequent stages of our study.

**XGBoost**

By feeding the tumor tiles into the four binary OvR classifiers, we obtained eight scores for each tile, two from each classifier. Each pair of scores represents the classifier's confidence in the images belonging to either the target class or the rest class. To aggregate the tile-wise predictions of four OvR CNNs to classify the molecular subtype of WSIs, we quantified the number of tiles within a WSI that had scores exceeding the optimal threshold for each class. This process involved counting the tiles classified as LumA, ¬LumA (not-LumA), LumB, ¬LumB, HER2, ¬HER2, BL, and ¬BL. Subsequently, these eight values were used as new features to train an XGBoost model, which predicts the molecular subtype of the WSIs (Fig. 3). XGBoost, an implementation of gradient-boosted decision trees, is a highly flexible and versatile machine learning tool, proven effective in a wide range of supervised learning tasks, demonstrating its ability to learn complex patterns in large volumes of data [27, 36].

To train and fine-tune the XGBoost model, we allocated 15% of the entire dataset (the yellow part of the pie chart in Fig. 2), which was further divided randomly into 80% for training and 20% for validation. We used the thresholds established in the earlier classification step to count the tiles predicted as either the target class or the rest class for all four breast cancer molecular subtypes in each WSI.

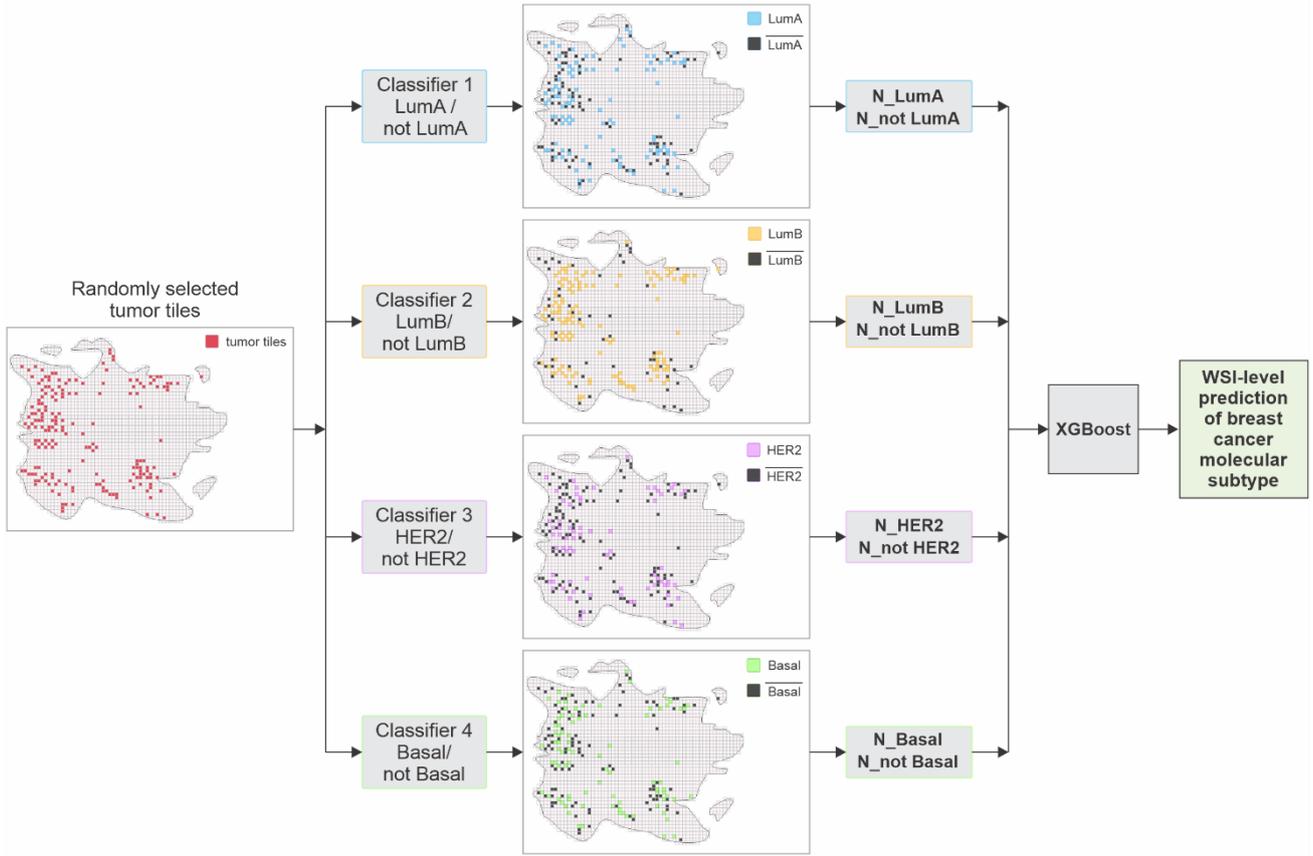

**Figure 3**: Aggregation of four OvR binary classifiers for predicting breast cancer molecular subtypes in a WSI. Randomly selected tumor tiles are independently processed by each classifier. The counts of tiles classified into target and non-target classes are used as features in an XGBoost model, which integrates the results from all four classifiers to determine the molecular subtype.

## Results

### Classification of tumor and non-tumor tiles

The model trained on 75,799 tiles labeled as 'tumor' or 'non-tumor' from TCGA-BRCA and BRACS datasets was evaluated using the F1 score, achieving a value of 0.954. The F1 score provides a balance between precision and recall (sensitivity), making it a suitable metric for our classification task. Such a high F1 score indicates that the model is reliable and performs very well in distinguishing between tumor and non-tumor tiles. Table 3 demonstrates additional performance metrics of our model on the test set. In addition, the model's overall accuracy was 0.955, indicating a high level of overall correct predictions across both classes (Fig. 4).

**Table 3:** Classifier's performance metrics on breast histopathology tumor/non-tumor tiles

| Class | F1 score | Precision | Sensitivity | Specificity | Accuracy |
| --- | --- | --- | --- | --- | --- |
| Tumor | 0.954 | 0.963 | 0.945 | 0.965 | 0.965 |

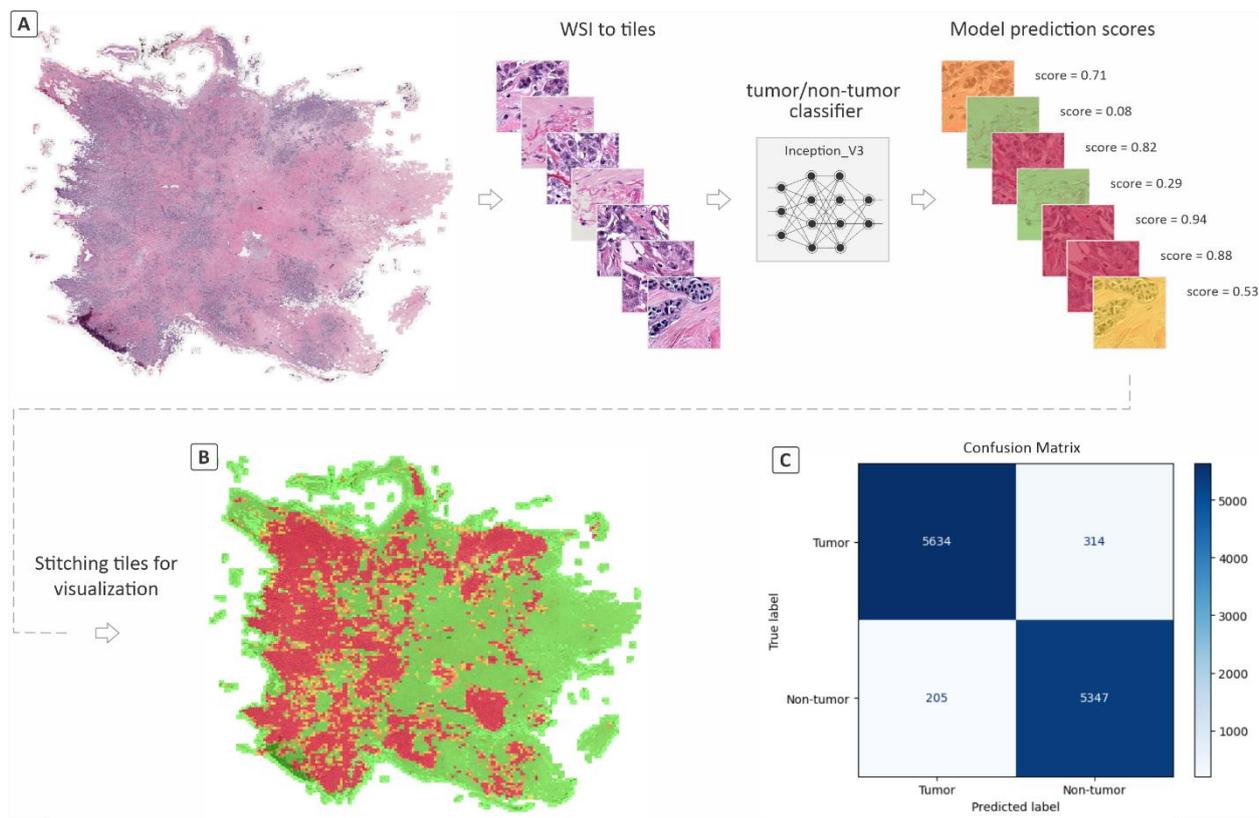

**Figure 4**: Performance of the classification model in distinguishing between tumor and non-tumor tiles. (A) Original WSI from the TCGA-BRCA dataset. (B) Recreated WSI by stitching the tiles with an overlay heatmap to illustrate tumor areas in the WSI. Red and yellow areas in the stitched image show tumor areas with high and low confidence scores, while green areas show non-tumor areas. (C) Confusion matrix showing the actual labels versus the model's predictions.

**Classification of breast cancer molecular subtypes**

We evaluated the entire pipeline on a hold-out test set (15% of the entire data, shown as the red part in the pie chart of Fig. 2) consisting of 221 H&E WSIs that had not been used in any part of the workflow. Table 4 presents the classification results for breast cancer molecular subtypes within the test set at the WSI level, obtained using an XGBoost model that aggregates predictions from four binary OvR models.

**Table 4:** WSI level classification metrics of breast cancer molecular subtypes with 95% confidence intervals

| Class | #WSIs | F1 score | Precision | Sensitivity | Specificity |
|---|---|---|---|---|---|
| LumA | 101 | 0.922 (0.880, 0.956) | 0.913 (0.856, 0.963) | 0.931 (0.873, 0.973) | 0.925 (0.876, 0.969) |
| LumB | 43 | 0.742 (0.629, 0.833) | 0.667 (0.537, 0.788) | 0.837 (0.714, 0.941) | 0.925 (0.876, 0.969) |
| HER2 | 32 | 0.545 (0.364, 0.690) | 0.652 (0.435, 0.840) | 0.469 (0.286, 0.647) | 0.925 (0.876, 0.969) |
| Basal | 45 | 0.698 (0.571, 0.800) | 0.732 (0.587, 0.868) | 0.667 (0.532, 0.792) | 0.925 (0.876, 0.969) |
| Macro-average | | 0.727 | 0.741 | 0.726 | 0.930 |

The performance of our model varies significantly across different breast cancer subtypes. We present the F1 score as our primary metric, commonly used in multi-class classification problems to balance precision and sensitivity, thereby providing a more comprehensive view of model performance. For the LumA subtype, the model achieved an F1 score of 0.922, indicating strong performance in correctly identifying cases of this subtype. Conversely, the HER2 subtype presented more of a challenge, with an F1 score of 0.545. This lower score suggests that while the model is particularly effective at ruling out false positives (high specificity), it frequently misses actual cases of HER2 (low sensitivity). For the LumB and BL subtypes, the F1 scores were 0.742 and 0.698, respectively, suggesting a moderate capability of the model to differentiate these subtypes. However, these scores also highlight potential limitations in distinguishing these subtypes from others or each other.

Given the imbalance in class distribution, with LumA comprising a significant majority of the WSIs in the test set (46%), to ensure a balanced evaluation, we employed a macro averaging approach to report the accuracy. Macro averaging calculates metrics independently for each class and then computes their average, thereby treating all classes with equal importance, irrespective of their frequency. The individual accuracy rates for LumA, LumB, HER2, and BL were 0.931, 0.837, 0.469, and 0.667, respectively, resulting in a macro average accuracy of 0.726.

Fig. 5 shows our classifier's confusion matrix and PR curves. The confusion matrix presents a detailed insight into the model's ability to classify breast cancer molecular subtypes at the WSI level. For LumA, the model demonstrates a high accuracy, correctly identifying 94 out of 101 instances, reflecting a solid capability to distinguish this subtype from others with minimal confusion, as indicated by the misclassification of only seven LumA cases as LumB (n=3) and HER2 (n=4). LumB WSIs were classified with lower accuracy; out of 43 samples, 36 were correctly identified. Most misclassifications occurred with the BL subtype (n=5), suggesting possible similarities in the features recognized by the model between these two subtypes. The classification of the HER2 subtype poses significant challenges, with only 13 correct predictions out of 32 samples, highlighting a critical area of weakness in the model's performance. Misclassifications are broadly distributed across all other subtypes, implying a fundamental difficulty in isolating defining features of HER2 within the model's current framework. BL subtype classification shows a better outcome, with 30 out of 45 cases correctly identified. However, misclassifications into LumB and HER2 suggest overlapping characteristics or insufficient specificity in the model's learning parameters.

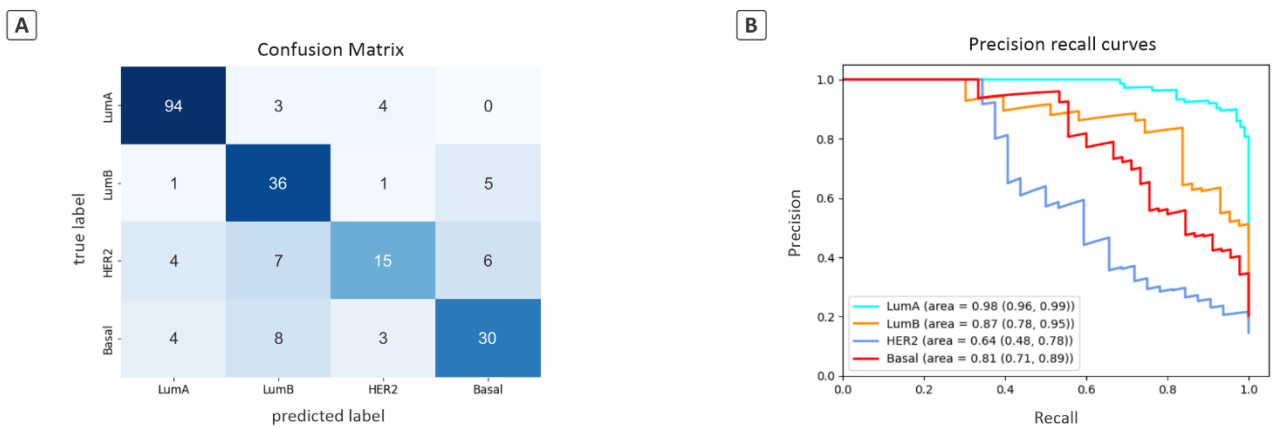

**Figure 5**: Performance of the XGBoost model in classifying breast cancer molecular subtypes. (A) Confusion matrix of model predictions. (B) PR curves of the subtypes with 95% confidence intervals.

Due to class imbalance in our dataset, we opted to use PR curves instead of receiver operating characteristic (ROC) curves. PR curves are more informative than ROC curves for imbalanced datasets as they focus on the classifier's performance with respect to the minority class, offering a clearer picture of the trade-offs between precision and recall. The PR curves with 95% confidence intervals for each breast cancer molecular subtype, presented in Fig. 5, provide additional insights into the model's performance variability.

The PR curve for the LumA subtype displayed an exemplary area under the curve (AUPRC) of 0.98, supported by a tight confidence interval ranging from 0.96 to 0.99. This result underscores the model's consistent and precise capability in identifying the LumA subtype. Conversely, the LumB subtype recorded an AUPRC of 0.87 with a broader confidence interval between 0.78 and 0.95, reflecting reliable yet slightly variable performance, which suggests minor challenges in accurate classification. The model exhibited more pronounced difficulties with the HER2 subtype, where the AUPRC was significantly lower at 0.64, and the confidence interval widened from 0.48 to 0.78, highlighting substantial inconsistency and overlap with other subtypes in its classification. Lastly, the Basal subtype attained an AUPRC of 0.81 and a confidence interval from 0.71 to 0.89, indicating a moderate level of performance with some degree of variability. This thorough analysis of the precision-recall curves and their respective confidence intervals builds upon the earlier discussed F1 scores and accuracy metrics to provide a detailed assessment of the model's classification effectiveness and pinpoint areas needing further improvement.

**Discussion**

In this study, we developed a supervised deep learning model to investigate whether H&E-stained histopathological images contain sufficient information for classifying breast cancer molecular subtypes. IHC staining is fundamental for molecular subtyping of breast cancer, offering greater precision than H&E staining but at a higher cost and longer processing time. However, it is susceptible to inter-observer variability, which can lead to diagnostic discrepancies. IHC-based classification may not always correspond with gene expression profiles, with discrepancies up to 31% [37]. Additionally, about one out of five LumB cases are identified as HER2 positive through IHC analysis, highlighting the complexity of differentiating between these subtypes [6]. Therefore, achieving diagnostic information from H&E WSIs comparable to that obtained through IHC would significantly reduce the time and costs associated with the diagnostic process, including both laboratory work and pathologists' evaluation.

To address this, we developed a two-step process. First, we trained a tile-wise CNN classifier to use only tumor regions for the classification of breast molecular subtypes. Our model achieved an overall F1 score and accuracy of 0.954 and 0.955, respectively, indicating a high effectiveness in identifying tumor-containing tiles. This high level of performance is critical for ensuring that subsequent analyses for subtype classification are conducted only on relevant tumor tissues.

The main part of the workflow focused on classifying breast cancer molecular subtypes, where we used the OvR strategy to simplify the problem of multi-class classification into a series of binary classification tasks, particularly given the complexity and similarity among the histology images of different breast cancer molecular subtypes. In this approach, the goal is to distinguish a specific class from all others. One benefit of this method is its interpretability; since each class is represented by a single classifier, it allows for a better understanding of the target class by analyzing its corresponding classifier. Moreover, the aggregation of predictions from four CNN binary classifiers with an XGBoost classifier model allowed us to leverage the strengths of binary classification while effectively addressing the complexities inherent in the classification of breast cancer molecular subtypes.

Our results revealed that the model performed differently across various subtypes with a macro F1 score of 0.72. It exhibited strong discrimination for LumA and LumB subtypes, but there were areas for improvement, especially for the HER2 class. The challenges in achieving high precision and sensitivity may be attributed to the underlying complexity of histopathological images, the variability in staining procedures, and the subtle differences that separate the molecular subtypes of breast cancer in practice.

While a comparison of several similar studies follows, it is important to acknowledge that using deep learning models for analyzing histopathological WSIs is a relatively young field of research. In addition, the existing studies have used different datasets, which limits the direct comparability of the findings. Despite this, we present performance metrics from various studies below to offer readers a general impression of how our results stack up against others in the field. Couture et al. [14] trained a VGG-16 binary classifier specifically for BL and non-BL tumors, achieving an overall accuracy of 0.77, with a sensitivity of 0.78 and specificity of 0.73, using a private dataset. In contrast, our model, which is not explicitly trained to distinguish between BL and non-BL tumors, can still be evaluated in this binary context by grouping LumA, LumB, and HER2 as non-BL. In this adjusted setting, our model achieves an overall accuracy of 0.75 for BL and non-BL WSIs with a sensitivity of 0.67 and a specificity of 0.94.

Jaber et al. [15] applied a deep learning model to extract features from multiscale tiles of WSIs in the TCGA-BRCA dataset. Their approach involved using Principal Component Analysis (PCA) alongside gene expression data to exclude patients with tumors exhibiting heterogeneous subtype characteristics from the training data. Using a multiclass support vector machine (SVM) algorithm, they achieved an accuracy of 0.66 for WSIs in the test set, which is lower than the macro average of our multi-class model (0.73). In addition, they trained a binary classifier for BL/non-BL tumors with an AUC-ROC of 0.86 and an accuracy of 0.87 for a balanced dataset. Despite its advantages, their model still faced challenges, such as the risk of misclassification due to subtype heterogeneity and the non-cancer richness of some tiles within cancer-rich clusters.

Abbasi et al. [16] investigated the predictiveness of morphological features in H&E-stained tissues for the classification of breast molecular subtypes. They leveraged a specialized backbone pre-trained on H&E WSIs in a self-supervised setting for large unlabeled data. Their network showcased enhanced generalizability to unseen data from new scanner types despite the dataset's relatively modest size used for backbone pre-training. Their model achieved an average AUC-ROC of 0.73. Their results showed the inconsistent performance of two models with different backbones when tested on unseen data from different scanners, which points to the potential of model tuning specific to the imaging source.

Liu et al. [17] proposed a weakly supervised learning model, where they used co-teaching to reduce the effect of noise patches and multi-instance learning for the classification of WSIs. Their top multi-class classifier demonstrated an accuracy of 0.58 and a macro F1-score of 0.65; however, aggregating the results of four binary classifiers and using the weighted fusion method improved their classification metrics to an accuracy and F1 of 0.64 and 0.69, respectively. They concluded that such AI models can be used for preliminary screening, making molecular subtyping more accessible and cost-effective, although it cannot yet replace traditional pathologist analysis entirely.

The XGBoost model in our study was trained on a dataset with imbalanced classes in WSI level. Training on imbalanced data, however, can lead to several issues, particularly affecting the model's performance for the less represented classes. Although we balanced the data at the tile level, this approach did not compensate for the imbalances at the WSI level, where the XGBoost model was trained. This imbalance was reflected in our results, with the model performing best for the major class (LumA) and underperforming for the minor class (HER2), which experienced the most misclassifications. Recognizing these challenges, future research could explore the impact of balancing the dataset on performance metrics. Specifically, adjusting the training set to better represent minority classes like HER2 could potentially improve the model's overall accuracy.

While our model demonstrated potential during internal validation (the hold-out test set), external validation is critical to ensure its generalizability and robustness across different datasets from diverse institutes and scanners. Our study lacked external test set validation due to the unavailability of publicly accessible datasets with the necessary annotations for breast cancer molecular subtypes. This limitation reflects a common challenge in this field, where high-quality, annotated data is scarce and often proprietary. External validation ensures that the model is not overfitted and performs reliably in practical applications. It also helps identify potential weaknesses when exposed to unseen data, highlighting areas that require further refinement. Demonstrating effectiveness across multiple institutions enhances credibility and facilitates broader acceptance within the research community. Additionally, the explainability of AI models remains an important consideration, as it provides insights into the decision-making process of the models. Although our work did not focus on explainability due to moderate model performance, future research should incorporate methods to enhance transparency and trust in the model's predictions.

## Conclusion

Our findings contribute to the evidence that H&E-stained WSIs contain pertinent information for classifying breast cancer molecular subtypes. Nonetheless, as this research is still emerging, our model, while promising, needs further enhancements and validation to ensure its generalizability. Future efforts could focus on refining these models to better manage the variability in histopathological images and extend their validation through external datasets.

## References

[1] Melina Arnold, Eileen Morgan, Harriet Rumgay, Allini Mafra, Deependra Singh, Mathieu Laversanne, Jerome Vignat, Julie R. Gralow, Fatima Cardoso, Sabine Siesling, and Isabelle Soerjomataram. Current and future burden of breast cancer: Global statistics for 2020 and 2040. The Breast, 66:15–23, 2022. ISSN 0960-9776. doi: https://doi.org/10.1016/j.breast.2022.08.010. URL https://www.sciencedirect.com/science/article/pii/S0960977622001448.

[2] Manzoor Ahmad Mir and Ifshana Mohi Ud Din. Molecular Subtypes of Breast Cancer and CDk Dysregulation, pages 133–148. Springer Nature Singapore, Singapore, 2023. ISBN 978-981-19-8911-7. doi: 10.1007/978-981-19-8911-7_6. URL https://doi.org/10.1007/978-981-19-8911-7_6.

[3] Diptendra Kumar Sarkar, editor. Breast Diseases, Guidelines for Management. CRC Press, 2024. ISBN 9780367421281 (hbk). doi: https://doi.org/10.1201/9780367821982.

[4] Gülin Alkan Şahin, Bedir Sümeyra Derin, Nahıda Valıkhanova, Betül Saraç, Ezgi Kacar, Nebi Serkan Demirci, Hulusi Fuat Demirelli, Ezgi Değerli, Nihan Şentürk Öztaş, and Hande Turna. Relationship between pathological response and molecular subtypes in locally advanced breast cancer patients receiving neoadjuvant chemotherapy. Journal of Chemotherapy, 35(1):29–38, 2023. doi: 10.1080/1120009X.2022.2043514. URL https://doi.org/10.1080/1120009X.2022.2043514. PMID: 35220928.

[5] Thi Minh Thuc Nguyen, Roanh Dinh Le, and Chu Van Nguyen. Breast cancer molecular subtype and relationship with clinicopathological profiles among Vietnamese women: A retrospective study. Pathology-Research and Practice, 250:154819, 2023. ISSN 0344-0338. doi: https://doi.org/10.1016/j.prp.2023.154819. URL https://www.sciencedirect.com/science/article/pii/S0344033823005198.

[6] Karen S Johnson, Emily F Conant, and Mary Scott Soo. Molecular Subtypes of Breast Cancer: A Review for Breast Radiologists. Journal of Breast Imaging, 3(1):12–24, 12 2020. ISSN 2631-6110. doi: 10.1093/jbi/wbaa110. URL https://doi.org/10.1093/jbi/wbaa110.

[7] Heung-Il Suk, Mingxia Liu, Xiaohuan Cao, and Jaeil Kim. Editorial: Advances in deep learning methods for medical image analysis. Frontiers in Radiology, 2, 2023. ISSN 2673-8740. doi: 10.3389/fradi.2022.1097533. URL https://www.frontiersin.org/articles/10.3389/fradi.2022.1097533.

[8] László Szilágyi and Levente Kovács. Special issue: Artificial intelligence technology in medical image analysis. Applied Sciences, 14(5), 2024. ISSN 2076-3417. doi: 10.3390/app14052180. URL https://www.mdpi.com/2076-3417/14/5/2180.

[9] Andrew Su, HoJoon Lee, Xiao Tan, Carlos J. Suarez, Noemi Andor, Quan Nguyen, and Hanlee P. Ji. A deep learning model for molecular label transfer that enables cancer cell identification from histopathology images. npj Precision Oncology, 6(1):14, 2022. ISSN 2397-768X. doi: 10.1038/s41698-022-00252-0. URL https://doi.org/10.1038/s41698-022-00252-0.

[10] Hossein Farahani, Jeffrey Boschman, David Farnell, Amirali Darbandsari, Allen Zhang, Pouya Ahmadvand, Steven J.M. Jones, David Huntsman, Martin Köbel, C. Blake Gilks, Naveena Singh, and Ali Bashashati. Deep learning-based histotype diagnosis of ovarian carcinoma whole-slide pathology images. Modern Pathology, 35(12):1983–1990, 2022. ISSN 0893-3952. doi: https://doi.org/10.1038/s41379-022-01146-z. URL https://www.sciencedirect.com/science/article/pii/S0893395222055107.

[11] Piumi Sandarenu, Ewan K. A. Millar, Yang Song, Lois Browne, Julia Beretov, Jodi Lynch, Peter H. Graham, Jitendra Jonnagaddala, Nicholas Hawkins, Junzhou Huang, and Erik Meijering. Survival prediction in triple negative breast cancer using multiple instance learning of histopathological images. Scientific Reports, 12(1):14527, 2022. ISSN 2045-2322. doi: 10.1038/s41598-022-18647-1. URL https://doi.org/10.1038/s41598-022-18647-1.

[12] R Rashmi, Keerthana Prasad, and Chethana Babu K Udupa. Breast histopathological image analysis using image processing techniques for diagnostic purposes: A methodological review. Journal of Medical Systems, 46(1):7, 2021. ISSN 1573-689X (Electronic), 0148-5598 (Print), 0148-5598 (Linking). doi: 10.1007/s10916-021-01786-9. URL https://pubmed.ncbi.nlm.nih.gov/34860316/.

[13] Kayvan Forouhesh Tehrani, Jaena Park, Eric J Chaney, Haohua Tu, and Stephen A Boppart. Nonlinear imaging histopathology: A pipeline to correlate gold-standard hematoxylin and eosin staining with modern nonlinear microscopy. IEEE Journal of Selected Topics in Quantum Electronics, 29(4 Biophotonics):6800608, 2023. ISSN 1077-260X (Print), 1558-4542 (Electronic), 1077-260X (Linking). doi: 10.1109/jstqe.2022.3233523. URL https://pubmed.ncbi.nlm.nih.gov/37193134/.

[14] Heather D Couture, Lindsay A Williams, Joseph Geradts, Sarah J Nyante, Ebonee N Butler, J S Marron, Charles M Perou, Melissa A Troester, and Marc Niethammer. Image analysis with deep learning to predict breast cancer grade, er status, histologic subtype, and intrinsic subtype. NPJ Breast Cancer, 4:30, 2018. ISSN 2374-4677. doi: 10.1038/s41523-018-0079-1.

[15] Mustafa I. Jaber, Bing Song, Clive Taylor, Charles J. Vaske, Stephen C. Benz, Shahrooz Rabizadeh, Patrick Soon-Shiong, and Christopher W. Szeto. A deep learning image-based intrinsic molecular subtype classifier of breast tumors reveals tumor heterogeneity that may affect survival. Breast Cancer Research, 22(1):12, 2020. ISSN 1465-542X. doi: 10.1186/s13058-020-1248-3. URL https://doi.org/10.1186/s13058-020-1248-3.

[16] Samaneh Abbasi-Sureshjani, Anıl Yüce, Simon Schönenberger, Maris Skujevskis, Uwe Schalles, Fabien Gaire, and Konstanty Korski. Molecular subtype prediction for breast cancer using h&e specialized backbone. In MICCAI Workshop on Computational Pathology, pages 1–9. PMLR, 2021.

[17] Hong Liu, Wen-Dong Xu, Zi-Hao Shang, Xiang-Dong Wang, Hai-Yan Zhou, Ke-Wen Ma, Huan Zhou, Jia-Lin Qi, Jia-Rui Jiang, Li-Lan Tan, Hui-Min Zeng, Hui-Juan Cai, Kuan-Song Wang, and Yue-Liang Qian. Breast cancer molecular subtype prediction on pathological images with discriminative patch selection and multi-instance learning. Frontiers in Oncology, 12:858453, 2022. doi: 10.3389/fonc.2022.858453.

[18] Sophie Foersch, Carsten Glasner, Ann-Christin Woerl, et al. Multistain deep learning for prediction of prognosis and therapy response in colorectal cancer. Nature Medicine, 29:430–439, 2023. doi: 10.1038/s41591-022-02134-1. URL https://doi.org/10.1038/s41591-022-02134-1.

[19] The Cancer Genome Atlas (TCGA). Genomic Data Commons Data Portal (GDC). https://portal.gdc.cancer.gov/projects/TCGA-BRCA. Accessed 07 Jul. 2023.

[20] BRACS: BReAst Carcinoma Subtyping. Institute of High-Performance Computing and Networking. 2020. https://www.bracs.icar.cnr.it/. Accessed 07 Jul. 2023.

[21] Angel Cruz-Roa, Hannah Gilmore, Ajay Basavanhally, Michael Feldman, Shridar Ganesan, Natalie Shih, John Tomaszewski, Anant Madabhushi, and Fabio González. High-throughput adaptive sampling for whole-slide histopathology image analysis (hashi) via convolutional neural networks: Application to invasive breast cancer detection. PloS one, 13(5), 2018. Accessed 07 Jul. 2023 via https://datadryad.org/stash/dataset/doi:10.5061/dryad.1g2nt41.

[22] Masoud Tafavvoghi, Lars Ailo Bongo, Nikita Shvetsov, Lill-Tove Rasmussen Busund, and Kajsa Møllersen. Publicly available datasets of breast histopathology he whole-slide images: A scoping review. Journal of Pathology Informatics, 15:100363, 2024. doi: 10.1016/j.jpi.2024.100363.

[23] National Cancer Institute Clinical Proteomic Tumor Analysis Consortium. The Clinical Proteomic Tumor Analysis Consortium Breast Invasive Carcinoma Collection (CPTAC-BRCA). The Cancer Imaging Archive; 2020. https://wiki.cancerimagingarchive.net/pages/viewpage.action?pageId=70227748. Accessed 07 Jul. 2023.

[24] Her2 Scoring Contest. Tissue Image Analytics (TIA) Centre. 2016. https://warwick.ac.uk/fac/cross_fac/tia/data/her2contest/. Accessed 07 Jul. 2023.

[25] Pete Bankhead et al. Qupath: Open source software for digital pathology image analysis. Available at https://qupath.github.io/, 2017. Version 0.1.2.

[26] Marc Macenko, Marc Niethammer, J. S. Marron, David Borland, John T. Woosley, Xiaojun Guan, Charles Schmitt, and Nancy E. Thomas. A method for normalizing histology slides for quantitative analysis. In 2009 IEEE International Symposium on Biomedical Imaging: From Nano to Macro, pages 1107–1110, 2009. doi: 10.1109/ISBI.2009.5193250.

[27] Tianqi Chen and Carlos Guestrin. Xgboost: A scalable tree boosting system. Proceedings of the 22nd ACM SIGKDD International Conference on Knowledge Discovery and Data Mining, 2016. URL https://api.semanticscholar.org/CorpusID:4650265.

[28] Christian Szegedy, Vincent Vanhoucke, Sergey Ioffe, Jon Shlens, and Zbigniew Wojna. Rethinking the inception architecture for computer vision. In Proceedings of the IEEE conference on computer vision and pattern recognition, pages 2818–2826, 2016.

[29] Juanying Xie, Ran Liu, Joseph Luttrell, and Chaoyang Zhang. Deep learning based analysis of histopathological images of breast cancer. Frontiers in Genetics, 10, 2019. ISSN 1664-8021. doi: 10.3389/fgene.2019.00080. URL https://www.frontiersin.org/journals/genetics/articles/10.3389/fgene.2019.00080.

[30] Songhui Diao, Weiren Luo, Jiaxin Hou, Ricardo Lambo, Hamas A. AL-kuhali, Hanqing Zhao, Yinli Tian, Yaoqin Xie, Nazar Zaki, and Wenjian Qin. Deep multi-magnification similarity learning for histopathological image classification. IEEE Journal of Biomedical and Health Informatics, 27(3):1535–1545, 2023. doi: 10.1109/JBHI.2023.3237137.

[31] Kaiming He, Xiangyu Zhang, Shaoqing Ren, and Jian Sun. Deep residual learning for image recognition. In 2016 IEEE Conference on Computer Vision and Pattern Recognition (CVPR), pages 770–778, 2016. doi: 10.1109/CVPR.2016.90.

[32] Yang Tan, Li juan Feng, Ying he Huang, Jia wen Xue, Li ling Long, and Zhen-Bo Feng. A comprehensive radiopathological nomogram for the prediction of pathological staging in gastric cancer using ct-derived and wsi-based features. Translational Oncology, 40:101864, 2024. ISSN 1936-5233. doi: https://doi.org/10.1016/j.tranon.2023.101864. URL https://www.sciencedirect.com/science/article/pii/S1936523323002504.

[33] Qingrong Sun, Weixiang Zhong, Jie Zhou, Chong Lai, Xiaodong Teng, and Maode Lai. Rcdpia: A renal carcinoma digital pathology image annotation dataset based on pathologists, 2024.

[34] Takaya Saito and Marc Rehmsmeier. The precision-recall plot is more informative than the roc plot when evaluating binary classifiers on imbalanced datasets. PloS one, 10(3), 2015. doi: 10.1371/journal.pone.0118432. URL https://doi.org/10.1371/journal.pone.0118432.

[35] Jason Brownlee. Roc curves and precision-recall curves for imbalanced classification, 2018. URL https://machinelearningmastery.com/roc-curves-and-precision-recall-curves-for-imbalanced-classification. Accessed: 2024-05-16.

[36] Xin Yu Liew, Nazia Hameed, and Jeremie Clos. An investigation of xgboost-based algorithm for breast cancer classification. Machine Learning with Applications, 6:100154, 2021. ISSN 2666-8270. doi: https://doi.org/10.1016/j.mlwa.2021.100154. URL https://www.sciencedirect.com/science/article/pii/S2666827021000773.

[37] Aleix Prat, Estela Pineda, Barbara Adamo, Patricia Galván, Aranzazu Fernández, Lydia Gaba, Marc Díez, Margarita Viladot, Ana Arance, and Montserrat Muñoz. Clinical implications of the intrinsic molecular subtypes of breast cancer. The Breast, 24–S35, 2015. ISSN 0960-9776. doi: https://doi.org/10.1016/j.breast.2015.07.008. URL https://www.sciencedirect.com/science/article/pii/S0960977615001460. 14th St.Gallen International Breast Cancer Conference–Proceedings Book.

# Appendix A. Supplementary material

**Supplement 1: Supplementary figures**

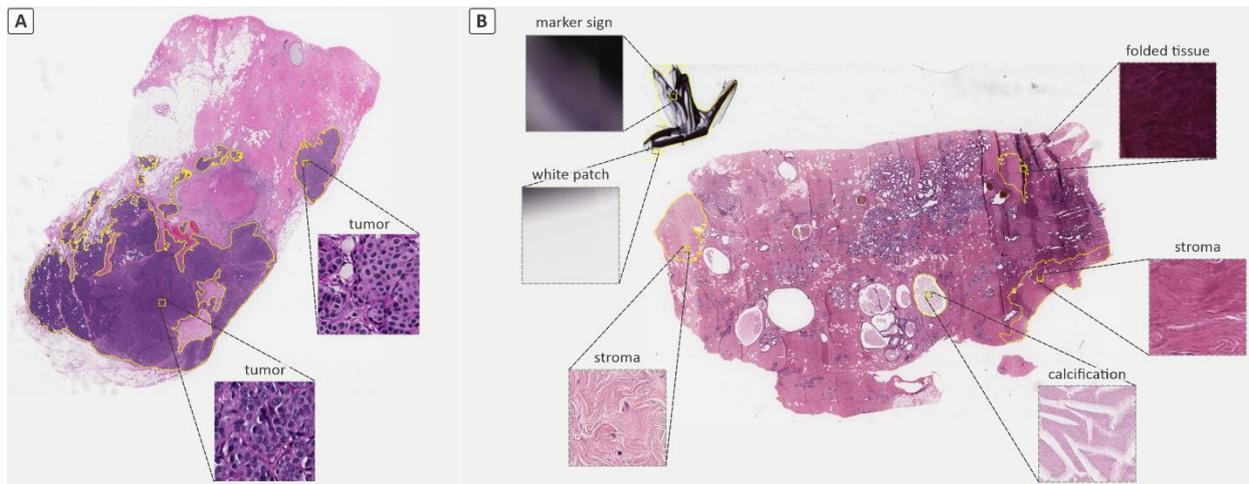

**Figure S1**: Extracting image tiles from breast H&E WSIs. (A) Tiling the annotated tumor regions from a WSI in the TCGA-BRCA dataset. (B) Examples of non-tumor tiles extracted from WSIs, including marker signs, normal tissues, folded tissue artifacts, and white areas of background.

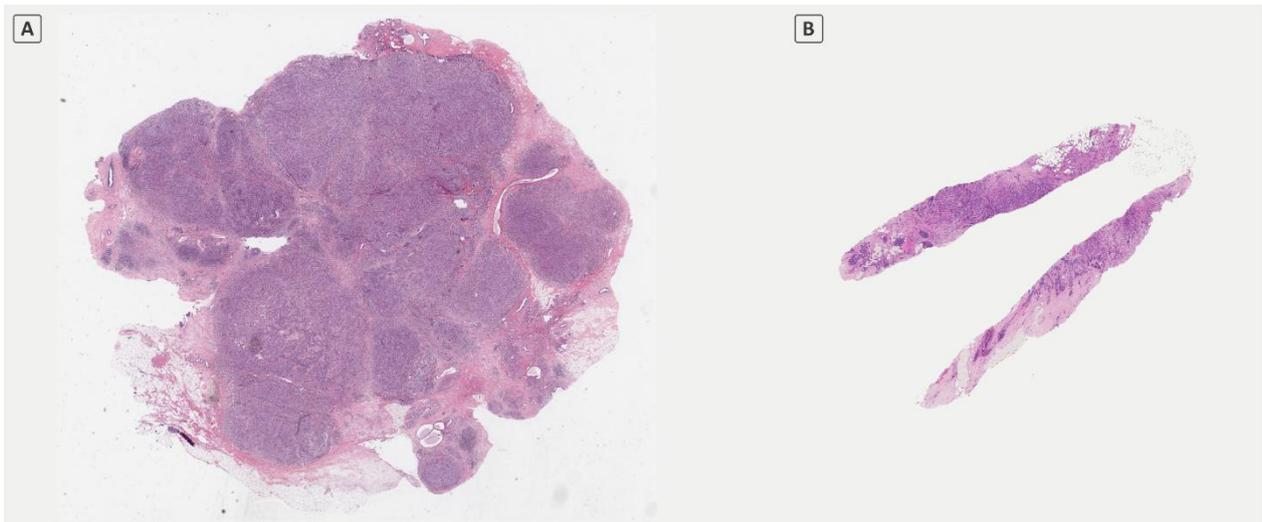

**Figure S2**: Comparison of harvested tissue between resection and biopsy: (A) a WSI from the TCGA-BRCA dataset, showing a breast tissue resection with an area of 2.32 cm$^2$, and (B) a WSI from the HER2-Warwick dataset, illustrating a breast biopsy with an area of 0.26 cm$^2$.

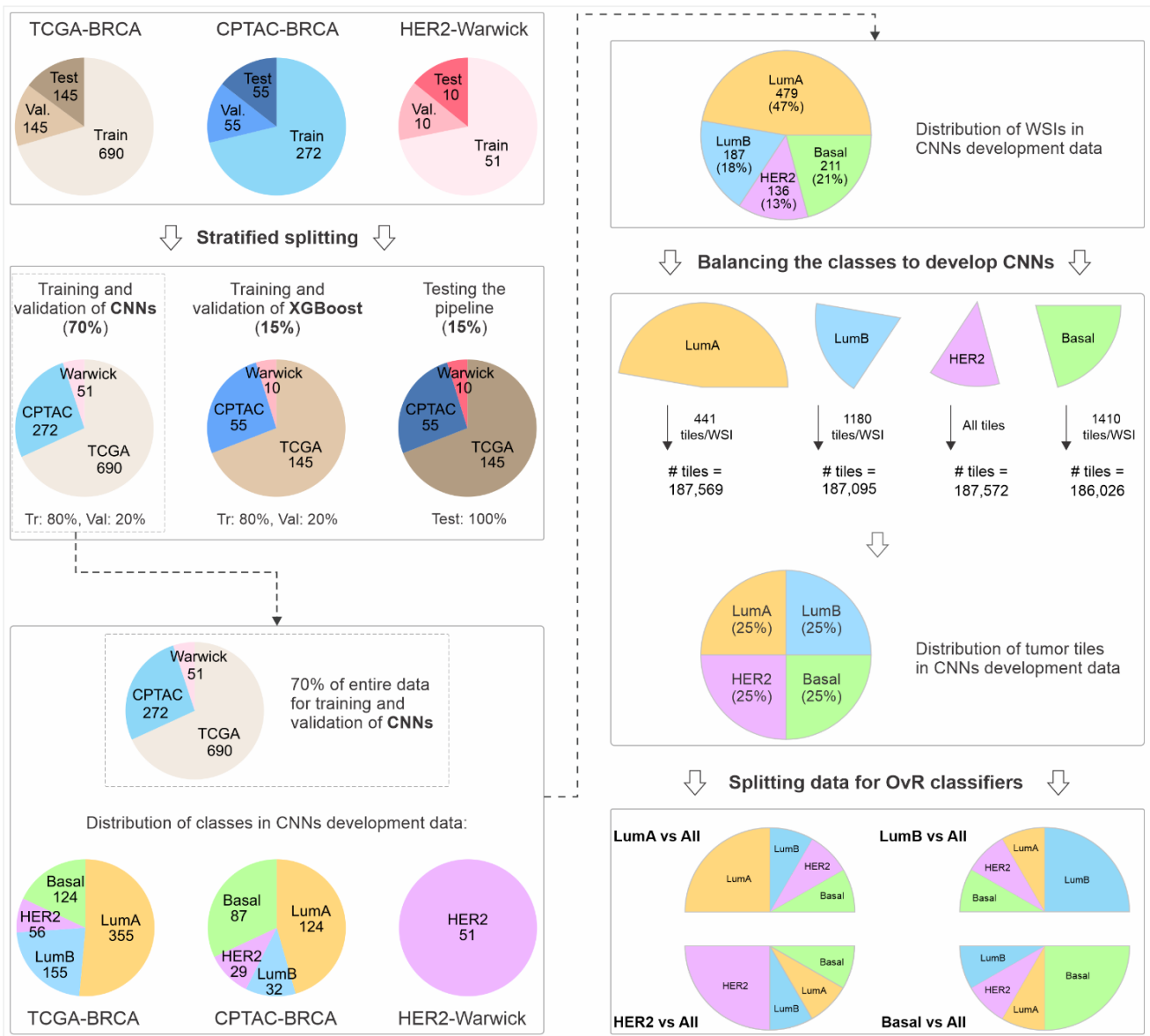

**Figure S3**: Illustration of data partitioning for training four binary One-vs-Rest classifiers.

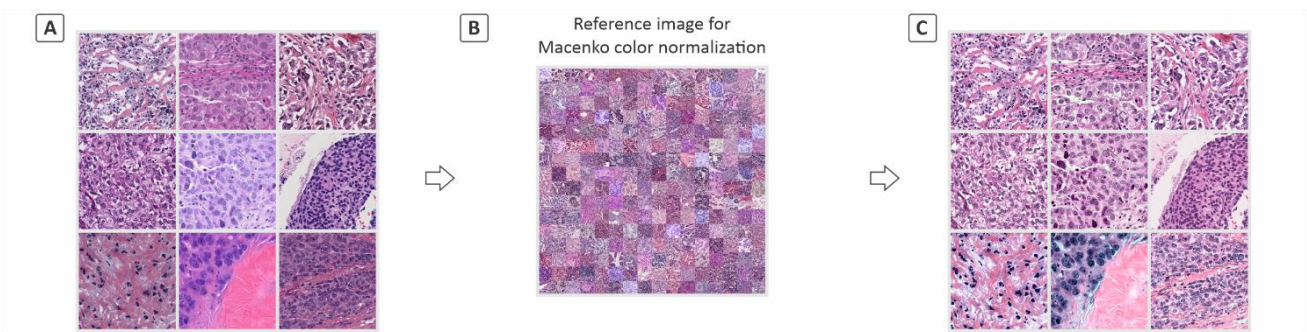

**Figure S4**: Color normalization of the tiles extracted from three different datasets: (A) Examples of original tiles showcasing inherent color variations in images. (B) The reference image, made out of 256 tiles from 256 randomly selected patients. (C) Normalized images demonstrating reduced color variation in image tiles.

16